\documentclass[%
 reprint,
nofootinbib,
 amsmath,amssymb,
 aps,prl
]{revtex4-1}

\usepackage{graphicx}
\usepackage{dcolumn}
\usepackage{bm}

\usepackage{amsmath,amssymb}
\usepackage{hyperref}
\usepackage{graphicx}
\usepackage{mathrsfs}
\usepackage{float}
\usepackage{ulem}
\usepackage{braket}
\usepackage{color}

 \newcommand{\bea}{\begin{eqnarray}}
\newcommand{\eea}{\end{eqnarray}}
\newcommand{\be}{\begin{equation}}
\newcommand{\ee}{\end{equation}}
\newcommand{\ba}{\begin{align}}
\newcommand{\ea}{\end{align}}

\newcommand{\sym}{\text{Sym}}

\newcommand\rref[1]{(\ref{#1})}

\begin{document}
\begin{flushright}
\hfill{\tt CERN-TH-2020-096}
\end{flushright}


\title{Random Statistics of OPE Coefficients and Euclidean Wormholes}

\author{Alexandre Belin$^{a}$, $\quad$
 Jan de Boer$^{b}$}

\affiliation{%
$^a$ CERN, Theory Division, 1 Esplanade des Particules, Geneva 23, CH-1211, Switzerland  \\
$^b$ Institute for Theoretical Physics, University of Amsterdam, PO Box 94485, 1090 GL Amsterdam, The Netherlands, 
}%

\begin{abstract}
We propose an ansatz for OPE coefficients in chaotic conformal field theories which generalizes the Eigenstate Thermalization Hypothesis and describes any OPE coefficient involving heavy operators as a random variable with a Gaussian distribution. In two dimensions this ansatz enables us to compute higher moments of the OPE coefficients and analyse two and four-point functions of OPE coefficients, which we relate to genus-2 partition functions and their squares. We compare the results of our ansatz to solutions of Einstein gravity in AdS$_3$, including a Euclidean wormhole that connects two genus-2 surfaces. Our ansatz reproduces the non-perturbative correction of the wormhole, giving it a physical interpretation in terms of OPE statistics. We propose that calculations performed within the semi-classical low-energy gravitational theory are only sensitive to the random nature of OPE coefficients, which explains the apparent lack of factorization in products of partition functions.
\end{abstract}

\pacs{Valid PACS appear here}
\maketitle

\section{Introduction}

Chaotic quantum many-body systems exhibit perhaps the strongest form of universality in physical systems, going from the random nature of the spectrum statistics (see e.g. \cite{Berry85,PhysRevB.55.1142}) to the fact that typical states thermalize (see \cite{Gogolin:2016hwy,DAlessio:2016rwt} for reviews). A quantitative formulation of this statement is known as the Eigenstate Thermalization Hypothesis (ETH) \cite{PhysRevA.43.2046,PhysRevE.50.888}
\be \label{ETH}
\braket{E_i | O_\alpha | E_j}=  \delta_{ij} f_\alpha(\bar{E}) + e^{-S(\bar{E})/2} g_\alpha(\bar{E},\delta E) R_{ij} \,,
\ee
where $O_\alpha$ is a simple (few-body) operator and $f$ and $g$ are smooth functions of the average energy $\bar{E}$ and energy difference $\delta E$. These functions encode the microcanonical expectation value of the one and two-point functions of $O$, respectively. The matrix $R_{ij}$ is a matrix of independent and identically distributed random variables with unit variance.

The intuition behind ETH is that simple operators cannot distinguish between energy eigenstates and that up to exponentially suppressed corrections in the entropy, their expectation values are given by a diagonal matrix made of the microcanonical expectation value. ETH gives a statistical interpretation to this correction as coming from a random matrix $R_{ij}$. In a definite quantum system with a fixed Hamiltonian, the numbers $R_{ij}$ will have definite values but one can nonetheless treat them as Gaussian random variables to good approximation.

In this letter, we will investigate a generalization of the ETH to very particular quantum systems: conformal field theories. Conformal field theories are characterized by two pieces of dynamical data: the spectrum of local operators (which are in one to one correspondence with the energy eigenstates $E_i$) and the OPE coefficients $C_{O_1O_2O_3}$ which dictate the fusion rules for these operators. It is important to distinguish two types of local operators: those which are very heavy and correspond to high energy states that we will denote by roman indices $O_i$, and those which are light which will be labeled by greek letters $O_\alpha$. The light operators should be viewed as simple and in conformal field theory, ETH is a statement about the OPE coefficients $C_{ij\alpha}$ \cite{Lashkari:2016vgj}.

There are other observables in conformal field theories that do not look like expectation values in definite states. An example which will be relevant for this work are higher-genus partition functions (or local correlation functions on higher genus surfaces). A higher genus observable will typically involve OPE coefficients $C_{ijk}$ where all three operators are heavy, which falls outside the regime of validity of the ETH. In this letter we propose a generalization of ETH for chaotic conformal field theories that captures the statistics of such observables. \footnote{By chaotic, we will mean any non-integrable CFT where the ETH ansatz applies.}  We propose the following ansatz
\begin{center}
{\bf OPE Randomness Hypothesis: }
\\
$C_{ijk},C_{ij\alpha},C_{i\alpha\beta}$ are random variables with (to leading approximation) a Gaussian distribution
\end{center}

Given this ansatz, we can compute any observable $O$ which is constructed from OPE coefficients. For $O_{ij\alpha}=C_{ij\alpha}$, this is nothing else than the ETH. However, our ansatz generalizes to arbitrary combinations of the $C$ which can be used to build other observables. Note that an extension to ETH in CFT$_2$ has already been advocated for in \cite{Collier:2019weq}, on the merits of asymptotic formulae for the averaged values of OPE coefficients, which relate to the variance of OPE coefficients. Our ansatz formalizes this statement.

The motivation of our proposal comes from the fact that high energy eigenstates cannot be easily distinguished. For expectation values of light operators, this is the driving force behind the ETH which states that energy eigenstates behave like thermal states to great approximation. Our proposal is that this feature extends beyond such observables, and that the fusion of three high-energy operators (or 2 light and one high-energy) must be a random variable as well. Note that unlike $C_{ij\alpha}$ which has a diagonal non-random piece, there is no such contribution for $C_{i\alpha\beta}$ or $C_{ijk}$.

In this letter, we will be interested in studying higher-point correlation functions of OPE coefficients. In general, we expect the consistency of the CFT to typically require the introduction of small non-Gaussianities, as we will discuss.\footnote{Note that this is in fact already necessary in the ETH. As stated in \rref{ETH}, the ansatz guarantees that the one and two-point functions of $O$ satisfy ETH. But the nature of the operator algebra, or said differently higher-point correlation functions of light operators must also satisfy ETH, which necessarily requires non-gaussianities. For example, they are crucial to correctly capture OTOCs in energy eigenstates \cite{Foini:2018sdb,Chan:2018fsp,Anous:2019yku,Murthy:2019fgs,Nayak:2019evx} (see also \cite{Dymarsky:2018ccu} for related discussions). Schematically, they take the form $R_{ij}^\alpha R_{kl}^\beta=\delta^{\alpha,\beta}\delta_{i,k}\delta_{j,l}+h(\Delta) C^{\alpha\beta}_\gamma R^{\gamma}_{ik}\delta_{j,l}+...$} The primary object of study will be the four-point statistics of heavy OPE coefficients
\be
O_{ijklmnopqrst} \equiv C_{ijk}\bar{C}_{lmn}C_{opq}\bar{C}_{rst} \,,
\ee
which we will calculate following our ansatz.

Our analysis is also closely related to the proposal of \cite{Pollack:2020gfa}, which studies the statistics of typical states obtained from Haar averaging over the micro-canonical energy window. While the spirit of their proposal is similar to ours, we propose an ansatz for the statistics of OPE coefficients which are more directly related to energy eigenstates, so we will not need to perform Haar averages. Moreover, the ansatz for OPE coefficient enables the computation of other observables than expectation values of light operators in heavy states. It would be nevertheless interesting to understand the Haar-average properties of OPE coefficients with only heavy operators.

\subsection{Gravity is a theory of random variables}

In this letter, we will mostly focus on holographic CFTs, namely maximally chaotic CFTs with a large number of degrees of freedom, such that the theories can be alternatively described by quantum gravity in asymptotically Anti-de Sitter space, following the AdS/CFT correspondence \cite{Maldacena:1997re}. The second aspect of our proposal, is that gravitational computations using semi-classical Einstein gravity are incapable of resolving the individual eigenstates, and therefore treat quantities like $C_{ijk}$ as a random variable $R_{ijk}$ with a Gaussian distribution (plus small corrections). In doing so, gravity makes an approximation, it makes a small error. We will show that this error explains puzzles related to the lack of factorization between partition functions in the AdS/CFT correspondence as put forward in \cite{Maldacena:2004rf} and recently revisited in \cite{Marolf:2020xie}.

In particular, we will propose an interpretation of the genus-2 Euclidean wormhole represented in Fig. \ref{wormhole} as a contribution that gravity picks up due to the random statistics of four OPE coefficients used to compute the square of a genus-2 partition function. This provides a statistical interpretation of the genus-2 wormhole.

Our proposal states that gravity can never distinguish the microscopic structure of OPE coefficients, and can at best compute a few smooth functions of the mean energy and energy differences related to variances or higher moments of the random variables.

Finally, we would like to comment on theories which require disorder averaging such as the SYK model or the dual of JT gravity, as well as their connections to Euclidean wormholes (see for example \cite{Saad:2019lba}). A disorder average introduces correlations between multiple disconnected copies of the quantum system, so Euclidean wormholes are no longer a puzzle in that circumstance. We believe it can be understood directly from our ansatz: while our ansatz should be viewed as a statistical approximation to the observables, it can become exact once the Hamiltonian is disorder-averaged. In such a setup, the product of two partition functions will then not factorize. The question of which averaging would need to be done in $\mathcal{N}=4$ SYM or the D1D5 CFT at strong coupling, in order to exactly randomize the OPE coefficients following our ansatz is a profound and interesting question that goes beyond the scope of this work. Nevertheless, we want to emphasize that the mechanism for this to happen can be logically embedded in our proposal.

\section{Statistics of OPE coefficients}

We will now study the square of OPE coefficients. For simplicity, we will focus on the case where all three operators have approximately the same energy such that $\Delta_{i,j,k}\gg |\Delta_i-\Delta_j|$ etc. We will also restrict to two-dimensional CFTs from now on. Our ansatz for such a quantity yields
\bea \label{OPEsqstats}
C_{ijk}\bar{C}_{lmn}&=& f(\Delta)\text{Sym}_{(ijk),(lmn)} [ \delta_{i,l}\delta_{j,m}\delta_{k,n} ]\\
&+& \sqrt{g(\Delta,\delta\Delta)}S_{ijklmn}  \notag \,,
\eea
where we have introduced $S$, a tensor with the appropriate symmetry properties between its indices. From the point of view of the random variables $R_{ijk}$, $S$ is a non-gaussianity which parametrizes possible corrections. The detailed structure of $S$ is currently unknown to us, but we must have
\bea
&\overline{S_{ijklmn}}=0   \,, 
\eea
where the overline notation will mean an average of each index centered at mean energies $\Delta_i,\Delta_j,...$ and over a sufficiently large energy band (typically the microcanonical window). In this paper, we will mostly work with all three indices centered around the same energy $\Delta_i=\Delta_j=\Delta_k=\Delta$. This gives
\be
\overline{C_{ijk}C_{lmn}^*}= f(\Delta)\text{Sym}_{(ijk),(lmn)} [ \delta_{i,l}\delta_{j,m}\delta_{k,n} ]   \,.
\ee

The function $f(\Delta)$ can be determined in the limit of asymptotically large energy for the operators $i,j,k$ by computing a genus-2 partition function and using modular invariance \cite{Cardy:2017qhl,Collier:2019weq}
\be \label{fdef}
f(\Delta) \approx \left(\frac{27}{16}\right)^{3\Delta} e^{-3\pi \sqrt{\frac{c}{3}\Delta}} \,,
\ee
where we have neglected some power-law corrections which will not be relevant for this work. We now turn to the variance of the square of OPE coefficients.

\subsection{Four OPE coefficients and the variance of $C_{ijk}\bar{C}_{ijk}$.}
Let us now compute the variance of the square of the OPE coefficients. Our ansatz gives
\bea
C_{ijk}\bar{C}_{lmn} C_{opq}\bar{C}_{rst}&=&f^2(\Delta) \text{Sym}[ \delta_{i,l}\delta_{j,m}\delta_{k,n} \delta_{o,r}\delta_{p,s}\delta_{qt}] \notag \\
&+&f(\Delta)\sqrt{g(\Delta)} \text{Sym}[ \delta_{i,l}\delta_{j,m}\delta_{k,n} S_{opqrst}] \notag \\
&+&g(\Delta)\sym[S_{ijklmn}S_{opqrst}^*] \notag \\
&+&\sqrt{h(\Delta,\delta \Delta)} T_{ijklmnopqrst} \,,
\eea
where the symmetrization is between indices $(ijk),(lmn),(opq),(rst)$ but also between pairs of three indices. $T$ is yet another tensor introduced to keep track of further non-gaussianities, but it will not play any role since it must average to zero. We can now consider the average over the energy band. We find
\bea \label{CCCC}
&\ &\overline{C_{ijk}\bar{C}_{lmn}C_{opq}\bar{C}_{rst}}=f^2(\Delta) \text{Sym}[ \delta_{i,l}\delta_{j,m}\delta_{k,n} \delta_{o,r}\delta_{p,s}\delta_{qt}] \notag \\
&\quad&\qquad \qquad\qquad+g(\Delta,\delta \Delta)\overline{\sym [S_{ijklmn}S_{opqrst}^*]}
\eea

The first term in fact contains two contribution that will be of interest to us. These are
\be \label{twocontributions}
f^2(\Delta) (\delta_{i,l}\delta_{j,m}\delta_{k,n} \delta_{o,r}\delta_{p,s}\delta_{qt}+\delta_{i,r}\delta_{j,s}\delta_{k,t} \delta_{l,r}\delta_{m,s}\delta_{nt})
\ee
corresponding to the two different Wick contractions of the four OPE coefficients (there is also a contraction $CC-\bar{C}\bar{C}$ which is of the same form as the second term). The appearance of these two terms has striking consequences, as it prohibits the factorization of genus-2 partition functions, as we will see in the following section.

\section{Genus-2 partition function from the CFT}

The main object of interest for this paper is the genus-2 partition function. A genus two-partition function is computed by a triple sum over states weighted by OPE coefficients
\be \label{genus2def}
Z_{g=2}=\sum_{i,j,k} C_{ijk} \bar{C}_{ijk} q_1^{\Delta_i}q_2^{\Delta_j} q_3^{\Delta_k} \,,
\ee
where $q_a=e^{2\pi i \tau_a}$ and $\tau_a$ are the moduli of the genus-2 surface parametrizing the length of three cycles, which can be thought of as three inverse "temperatures" (see for example \cite{Cardy:2017qhl} for the relation between $\tau_a$ and the period matrix). We will be interested in a particular slice of the moduli space where all three parameters are equal, namely
\be
\tau_{1}=\tau_2=\tau_3=\frac{i\beta}{2\pi} \,,
\ee
and we will study the "high-temperature" behaviour of the partition function, namely the limit $\beta\to0$. To evaluate the partition function, we need the high-energy behaviour of the OPE coefficients. This is precisely the function $f(\Delta)$ given in \rref{fdef}. From \cite{Cardy:2017qhl,Collier:2019weq}, we have
\be
\frac{C_{ijk} \bar{C}_{ijk}}{\rho(\Delta)^3}=f(\Delta) \approx \left(\frac{27}{16}\right)^{3\Delta} e^{-3\pi\sqrt{\frac{c}{3}\Delta}} \,,
\ee
where we have taken the three-energies to be the same and we wrote the large $c$ version of the equation. This equation is valid for Virasoro-primary operators. To compute the partition function, we also need to sum over descendants. This can be achieved by the means of adding a genus-2 block $\mathcal{F}$ to the partition function, whose contribution gives $\mathcal{F}\approx \left(\frac{27}{16}\right)^{-3\Delta}$. We thus have
\be
Z_{g=2}=\sum_{\Delta} e^{3\pi\sqrt{\frac{c}{3}\Delta}} e^{-3\beta \Delta} \,.
\ee
This expression can be evaluate by saddle-point, whose saddle reads
\be \label{saddleDelta}
\Delta^{*}= \frac{c}{12}\frac{\pi^2}{\beta^2} \,,
\ee
and we obtain
\be \label{CFTgenus2}
Z_{g=2} \approx e^{\frac{c}{4} \frac{\pi^2}{\beta}} \,.
\ee
\subsection{Square of the partition function}

We are now interested in computing the square of the partition function. Obviously, in a definite CFT the answer must simply be the square of \rref{CFTgenus2}, so any microscopic calculation would produce
\be
Z^{\rm microscopic}_{g=2\times g=2} \approx e^{\frac{c}{2} \frac{\pi^2}{\beta}} \,.
\ee
However, this is not the outcome of the calculation following our ansatz for randomness. The issue is the second contribution in \rref{twocontributions}. The first contribution gives indeed $Z^{\rm microscopic}_{g=2\times g=2}$, but there is a correction, given by
\be
\sum_{\Delta} e^{6\pi \sqrt{\frac{c}{3}\Delta} }e^{-3S(\Delta)} e^{-6\beta \Delta} \,,
\ee
where the factor of $e^{-3S(\Delta)}$ arises because of the nature of the index contractions for the second term. Using the Cardy formula \cite{cardyformula}, we find that there is no growing exponential and we simply have $\sum_{\Delta}  e^{-3\beta \Delta} $, which has no saddle-point and gives an $\mathcal{O}(1)$ answer. We thus find
\be \label{CFTZsq}
Z^{\rm ansatz}_{g=2\times g=2} \approx e^{\frac{c}{2} \frac{\pi^2}{\beta}}+\mathcal{O}(1) \,.
\ee
It is useful to define the connected part of the square of genus-2 partition functions
\be \label{CFTZsq}
Z^{\rm connected}_{g=2\times g=2} \equiv  Z_{g=2\times g=2} -(Z_{g=2})^2 \,.
\ee
Computing this quantity with our ansatz and using equation \rref{saddleDelta}, we find
\be
-\log \left[\frac{Z^{\rm connected}_{g=2\times g=2} }{(Z_{g=2})^2}\right] = \frac{c}{2}\frac{\pi^2}{\beta}=\frac{3}{2}S(\Delta) \,.
\ee

We will see in the following section that gravity reproduces the form of this answer, and this connected contribution is attributed to a Euclidean wormhole.

It is worth emphasizing two points. This connected contribution would prevent factorization of the square of partition functions had we picked different moduli for the two genus-2 surfaces: this is problematic, in that it does not agree with a microscopic calculation where the square of the partition function manifestly factorizes. This will have a direct resonance with the gravitational computation.

So far, we have not taken into account the contribution of the non-gaussianity $S$ in \rref{CCCC}. We currently do not have a concrete proposal for how to capture the precise structure of the tensor $S$, directly from the CFT. It is possible that it is itself a tensor of iid random variables, or it could have further substructure. It is certainly constrained by modular invariance of the genus-3 partition function, which also involves four OPE coefficients with some cyclic contractions. This is in spirit with ETH, certain contractions of the non-gaussianities are captured by consistency of the higher-point correlation functions, which for heavy operators naturally maps to higher-genus surfaces. \footnote{Unfortunately, not much is known about the scaling of higher genus partition functions.  We expect the scaling of the non-gaussianities of $C_{ijk}$ to be of order $e^{- G(g)S}$, for an increasing function $G(g)$ which is currently unknown. Similar observations have been made for the ETH in \cite{Foini:2018sdb}. It would be interesting to investigate whether this can be made precise.}

There also exists contractions of indices which are not of the cyclic type, and do not immediately seem constrained from the CFT. Turning the problem around, these structures could be extracted from the bulk, as we will see below: one could infer the properties of $S$ for holographic CFTs by going beyond the leading on-shell action of the wormhole and computing one-loop determinants, both for the handle-body and the wormhole. We expect the result to be less universal since it could depend on the matter content of the bulk theory. We hope to return to this question in the future.

\section{Genus-2 partition function in AdS$_3$ gravity}

We have proposed certain statistical properties \rref{OPEsqstats} of OPE coefficients in chaotic CFTs, and moreover that the dual gravitational description can only capture the random-matrix nature of the OPE coefficients. In this section, we provide some evidence for these proposal by direct calculations in AdS$_3$/CFT$_2$.

We will study three-dimensional gravity with a negative cosmological constant, whose action is given by
\be
S_{\rm grav}=-\frac{1}{16\pi G_N} \int d^x \sqrt{g} \left(R+\frac{2}{\ell^2}\right) \,,
\ee
where the AdS radius $\ell$ in Planck units is related to the central charge of the dual CFT \cite{Brown:1986nw}
\be
c= \frac{3\ell}{2G_N} \,.
\ee

The object we want to compute in gravity is a genus-2 partition function \rref{genus2def} (and the square thereof). By the standard AdS/CFT dictionary, the general prescription to compute such quantities is to find a solution to the gravitational equations of motion with the appropriate boundary conditions, in this case one (or two) genus-2 surfaces and evaluate
\be
Z_{g=2}\approx e^{-S_E^{\rm on-shell}}
\ee
where  $S_E^{\rm on-shell}$ is the (regularized) Euclidean action of the bulk solution.
 
\begin{figure}[h!]
\centering
\includegraphics[width=0.45\textwidth]{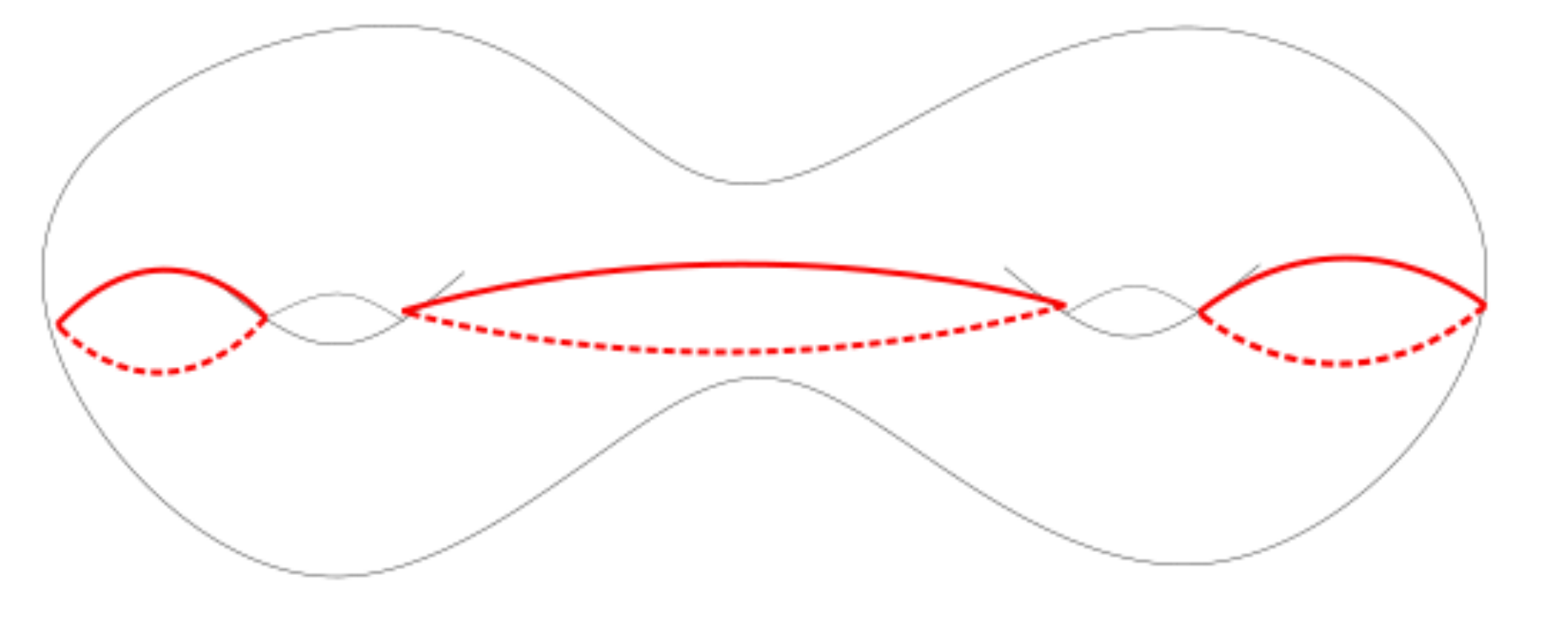}
\caption{A genus-2 surface with 3 cycles highlighted. One handlebody solution corresponds to a solid filing of the genus-2 surface such that the three red cycles are made contractible in the bulk.}
\label{goodchannel}
\end{figure}

The simplest solutions in gravity are called \textit{handlebody} solutions, which are quotiens of Euclidean AdS$_3$ corresponding to fillings of the genus-2 surface \cite{Krasnov:2000zq}. This is illustrated in Fig. \ref{goodchannel}. There are always multiple handlebodies for a given genus-2 surface, given by choices of cycles that are made contractible in the bulk. This is the generalization of thermal AdS and the BTZ black hole to genus-2 boundaries. Evaluating the action of the handlebody geometries is a complicated task, which boils down to the evaluation of a Liouville action \cite{Maxfield:2016mwh}. General results can only be obtained numerically, but in certain limits there are analytic expressions.

\begin{figure}[h!]
\centering
\includegraphics[width=0.25\textwidth]{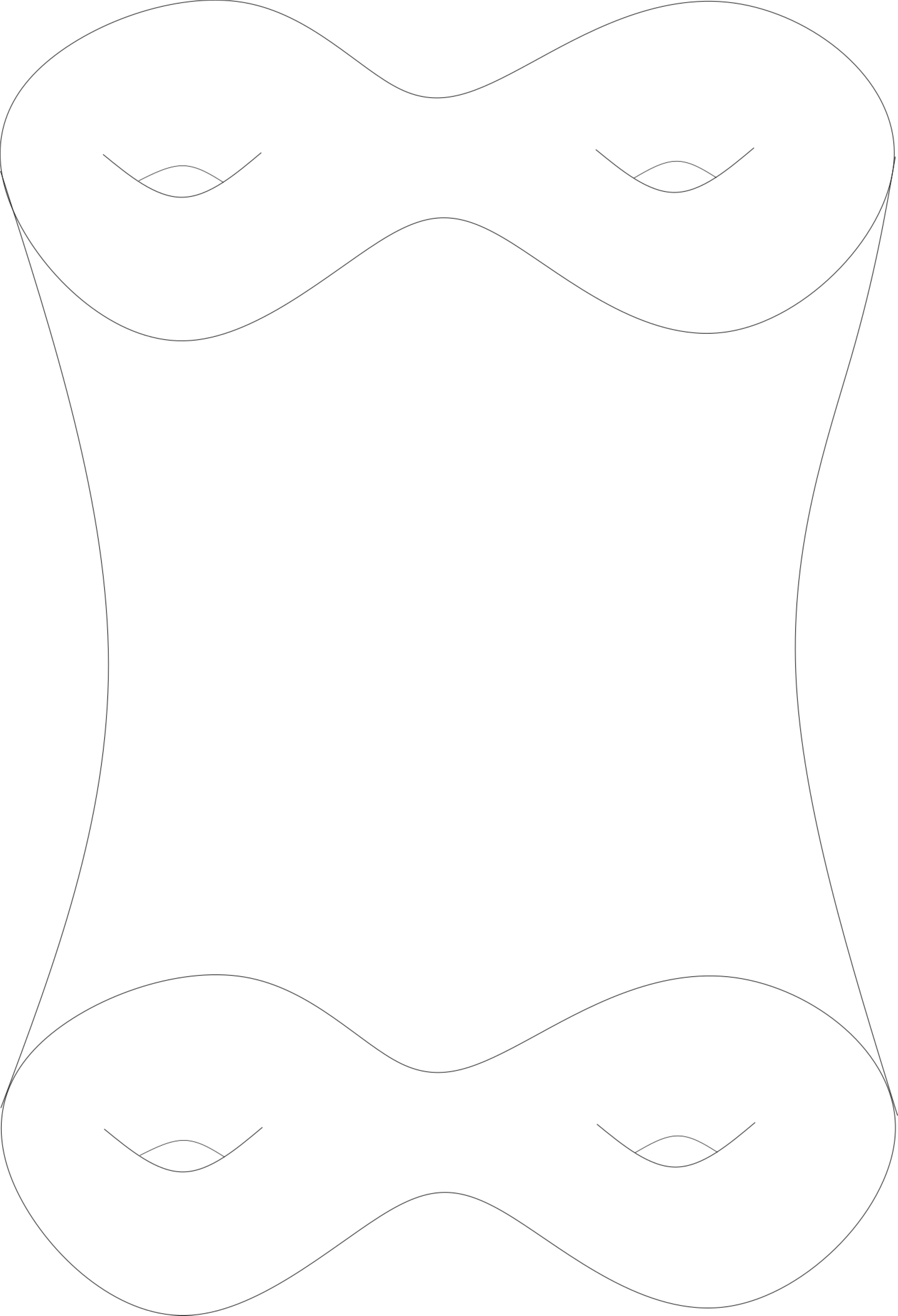}
\caption{The euclidean wormhole \rref{wormholemetric} connecting through the bulk two genus-2 surfaces that were originally disconnected.}
\label{wormhole}
\end{figure}

For a constant curvature metric on the genus-2 surface and in the limit where the three moduli (taken to be equal) degenerate $\tilde{\beta}_i=2\pi\tilde{\tau}_i\to0$, it can be shown \cite{Maxfield:2016mwh} that\footnote{We write $\tilde{\beta}$ for the modulus because the CFT answer \rref{CFTgenus2} was not written for the constant curvature metric and we must be careful in comparing moduli. We will do so shortly.}
\be \label{Sonshell}
S^{\text{h}}_{\text{on-shell}}= -  \frac{c}{2} \frac{\pi^2}{\tilde{\beta}} \,,
\ee
If we wish to compute the partition function for multiple disconnected CFTs, we can simply fill each CFT with a handlebody geometry. For example, we will be interested in computing the partition function for two identical but disconnected genus-2 surfaces. In this case, the spacetime is disconnected and the action of the two handlebodies add. We thus obtain
\be
Z^h_{g=2 \times g=2} \approx e^{c \frac{\pi^2}{\tilde{\beta}} } \,.
\ee
For multiple disconnected boundaries, there are also \textit{non-handlebody} solutions that connect the asymptotic boundaries through the bulk. The simplest and perhaps most famous example is the genus-2 wormhole \cite{Maldacena:2004rf}, whose metric is given by
\be \label{wormholemetric}
ds^2=\ell_{\rm AdS}^2 (d\tau^2+\cosh^2\tau d\Sigma_g^2) \,,
\ee
where $d\Sigma_g^2$ is the constant curvature metric on a genus-2 surface. This wormhole is represented in Fig. \ref{wormhole}. There exist other more complicated wormholes connecting two genus-2 surfaces with different moduli that we will not discuss here, but would be relevant for the statistics of OPE coefficients when the operators do not have the same mean energy.

The action for the symmetric wormhole is particularly easy to compute for any choice of moduli for the genus-2 surface, given the factorized form of the metric. The action is simply a constant, which can be chosen to be zero within a certain regularization scheme for the conformal anomaly \cite{Maxfield:2016mwh} and we will follow this convention. We will shortly consider ratios of partition functions where this constant is anyway meaningless. Now consider the following quantity:
\be
\frac{Z_{g=2\times g=2}}{(Z_{g=2})^2} \,.
\ee
The numerator can be computed given the two saddle-point solutions we have described, namely the two disconnected handlebodies and the euclidean wormhole. The denominator is the square of the handlebody answer. In total we have
\be
\frac{Z_{g=2\times g=2}}{(Z_{g=2})^2} \approx \frac{e^{c \frac{\pi^2}{\tilde{\beta}} } +1}{e^{c \frac{\pi^2}{\tilde{\beta}} }}  = 1+ e^{-c \frac{\pi^2}{\tilde{\beta}} } \,.
\ee
To compare to the CFT calculation, we must be careful to compare the moduli in the appropriate fashion. The CFT computation \rref{genus2def} was performed with a different choice of metric for the genus-2 surface, corresponding to a branched-cover of the plane, rather than a constant curvature metric.  The parameters labelling the size of the cycles are not equal, and we have $\tilde{\beta}=2\beta$. Taking this into account, our gravitational contribution precisely matches the square of the genus-2 partition function computed in \rref{CFTZsq} thanks to our ansatz of randomness for the OPE coefficients.

Finally, we can rewrite this expression in terms of the saddle-point energy corresponding to $\beta$ \rref{saddleDelta}. We find
\be
\frac{Z_{g=2\times g=2}}{(Z_{g=2})^2} \approx  1+ e^{-3\pi \sqrt{ \frac{c}{3}\Delta}}\,.
\ee
We now use the Cardy formula to give the answer in terms of the entropy, and we find
\be \label{finalresult}
\frac{Z_{g=2\times g=2}}{(Z_{g=2})^2} \approx 1+ e^{-\frac{3}{2}S(\Delta)} \,.
\ee

Note that the canonical and microcanonical variance differ, as the microcanonical variance would give a suppression of $e^{-3S}$. It is useful to consider the connected part of the gravitational calculation which only keeps the wormhole contribution. We find
\be
-\log \left[ \frac{Z_{g=2\times g=2}\big|_{\text{connected}}}{(Z_{g=2})^2} \right] = \frac{3}{2} S(\Delta) \,.
\ee
This can be matched with the connected contribution to the CFT answer obtained from our ansatz, following \rref{CFTZsq}. The factor of $3 S(\Delta)/2$ is unambiguously matched.

\subsection{Other OPE channels and handlebodies}

\begin{figure}[h!]
\centering
\includegraphics[width=0.45\textwidth]{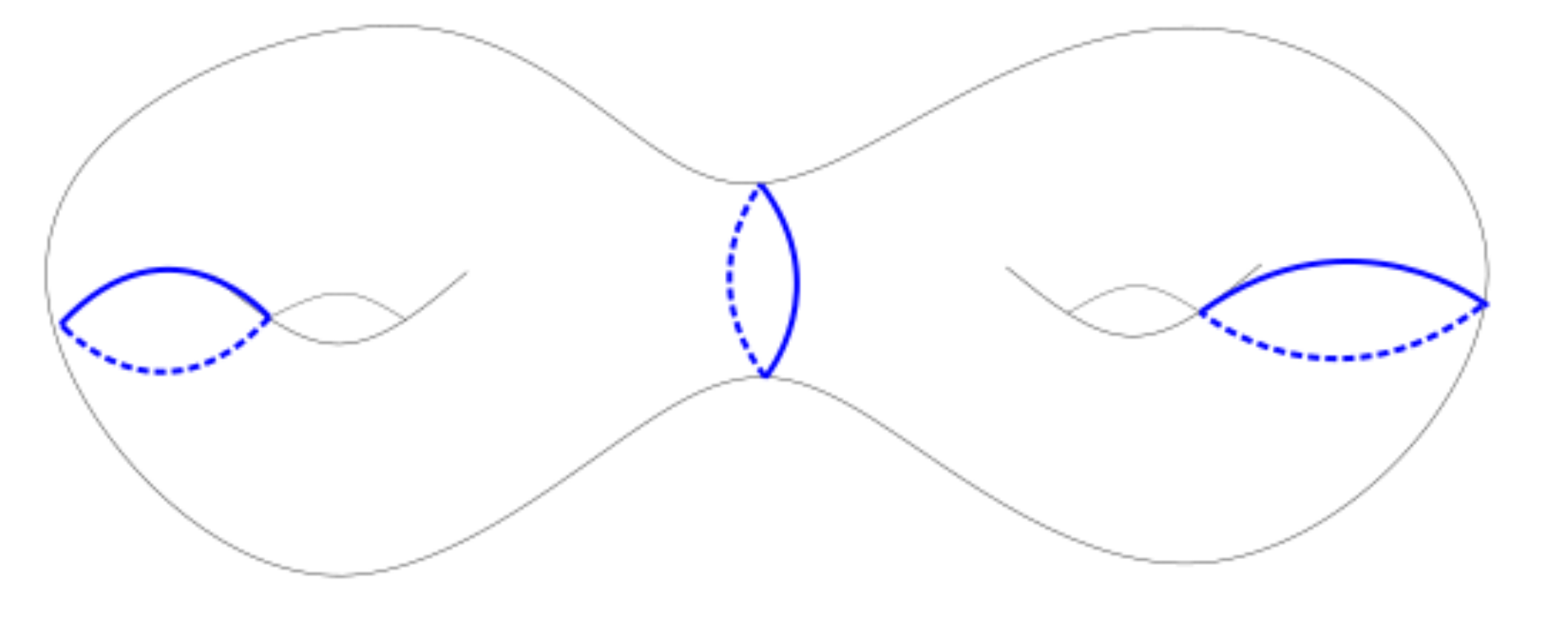}
\caption{A handlebody where the three blue cycles are made contractible in the bulk. This saddle is equivalent to decomposing the genus two partition function such that the three blue cycles are cut open. This yields a sum over OPE coefficients $\sum_{i,j,k}C_{iij}C_{jkk}$, corresponding to a product of torus one-point functions.}
\label{badchannel}
\end{figure}

A striking feature of the conjecture for the statistics of OPE coefficients is that one can perform different contractions of the indices given the ansatz. We will now argue that this is in fact natural from the bulk perspective, and corresponds to other handlebody geometries. The OPE channel can be directly connected to the cycles that are made contractible in the bulk. For example, a handlebody as in Fig. \rref{badchannel} corresponds to different bulk cycles made contractible.

We thus see that some of the other handlebody solutions are naturally encoded in the OPE ansatz. Note that for a given choice of a moduli, there will always be one dominant saddle, and thus the subleading saddles will be exponentially suppressed. This can be seen already in the microcanonical average over OPE coefficients \rref{OPEsqstats} where different index combinations are exponentially suppressed in the entropy, once contracted to compute the genus-2 partition function.

\section{Discussion}

In this letter, we have proposed an ansatz for the statistics of OPE coefficients in chaotic conformal field theories, identifying OPE coefficients of heavy operators as random variables. For large $c$ maximally chaotic theories, we compared our ansatz to calculations in gravity and showed that the first correction due to contractions of the random variables reproduces the contribution of a Euclidean wormhole connecting two boundary CFTs. We argued that gravitational computations using the low-energy effective theory only captures the random nature of OPE coefficients, which explains the lack of factorization. We will now discuss several open questions.

\subsection{Beyond the single large energy limit}

In this paper, we have focused our attention on the simplest possible kinematic regime: the case where all three operators of the OPE coefficient have the same (mean) energy which is taken to be parametrically large. One can also work out the details of the cases where all three energies are taken to be distinct (but all large). From the bulk, one would expect to get a connected component suppressed by $e^{-\frac{S(\Delta_1)+S(\Delta_2)+S(\Delta_3)}{2}}$, which again is in agreement with our ansatz.

A more subtle question to ask is the fate of our ansatz outside of the Cardy regime. The main challenge is that unlike the torus partition function, the genus-2 partition functions are no longer universal at large $c$ outside of the Cardy limit \cite{Belin:2017nze} (see \cite{Michel:2019vnk} for a related discussion of $C_{iab}$). From the bulk point of view, the handlebody geometries may no longer be the dominant saddle-point and solutions with a condensed scalar field can dominate \cite{Dong:2018esp}. It would be interesting to understand whether a connected wormhole with a condensed scalar field exists or not.

Finally, note there exists Euclidean wormholes with two genus-2 surfaces of different moduli. It would be interesting to try to probe the nature of the correlation between different energies in the OPE coefficient thanks to an analytical or numerical analysis of the asymmetric wormholes.

\subsection{Higher dimensions}

It would also be interesting to understand the extent to which our ansatz holds in higher dimensions. We believe that the structure of the OPE statistics is not vastly different, but perhaps the functions that enter in the ansatz are less universal than in $d=2$. For example, even the coefficient in the thermal entropy of CFTs on $S^1\times S^{d-1}$ is not universal, unlike the Cardy formula where it is given by the central charge due to modular invariance. Similarly, there is no formula for the asymptotics of OPE coefficients of three heavy operators in $d>2$ because there is no obvious counterpart of a genus-2 surface, i.e. a well-behaved geometry whose partition function probes OPE coefficients with three heavy operators (other asymptotic formulas for OPE coefficients with one or 2 heavy states do exist though \cite{Pappadopulo:2012jk,Delacretaz:2020nit}).

From the bulk point of view, there exists Euclidean wormholes in higher dimension, but the question of their stability is much more subtle (see for example \cite{Maldacena:2004rf}). In the simplest case, they are related to CFTs on $S^1 \times H^{d-1}/\Gamma$ where $\Gamma$ is a discrete subgroup of hyperbolic space. Since the spatial geometry has a negative curvature, the conformal coupling of the CFT fields can induce instabilities. Even if such wormholes were stable and thus valid saddle-points to the gravitational path-integral, it is not clear how to connect them to the local data of the theory. If anything, states on other topologies than the sphere are more likely to be connected to properties of non-local operators \cite{Belin:2018jtf}.

\subsection{Chaotic versus integrable theories}

The results we obtained were for large $c$ maximally chaotic CFTs dual to Einstein gravity in AdS$_3$. It is worth emphasizing that the Euclidean wormhole \rref{wormholemetric} is locally AdS and therefore is expected to be an $\alpha'$-exact solution to string theory. For example, one would expect it to be a valid saddle-point for the description of the D1D5 CFT at the orbifold point, where the theory is integrable. It would thus be tempting to write the answer \rref{finalresult} also for an integrable theory. How is this consistent with ETH which only holds in chaotic theories?

We are certainly not claiming that our ansatz for randomness is true in an integrable theory. In such a case, we expect there to be large deviations from the ansatz due to the many selection rules imposed by the infinite tower of conserved currents. Nevertheless, the genus-2 partition function only computes averages over the coefficients. It is thus possible that while the ansatz is not true for individual operators at the integrable point, it is still true on average. It would be interesting to understand this better. Finally, we would like to emphasize that our intuition for the connected contribution relies crucially on large $c$, where gravity is semi-classical. The evidence we have presented for our conjecture based on the Euclidean wormhole is only trustworthy at large central charge. It would of course be very interesting to understand the statistics of individual OPE coefficients for chaotic CFTs of small central charge. Unfortunately, only few theories of this type are known explicitly and it is not clear how to implement such a check, even numerically. Perhaps tensor networks of the like of MERA \cite{PhysRevLett.93.040502} would be able to probe some of these questions.

\subsection{Torus correlation functions and spectral form factor}

Note that our ansatz also predicts the existence of a connected contribution for torus one-point functions given by the statistics of
\be
\overline{C_{ii\alpha}\bar{C}_{jj\alpha}}=(f^{\alpha})^2(\Delta) + g^\alpha \delta_{i,j} \,.
\ee
Similar observations related to connected contribution have been made in JT gravity \cite{Saad:2019pqd}. Unfortunately, no wormhole solution of AdS$_3$ gravity exists with torus boundaries, leaving the explanation of this contribution as a puzzle.

We see several possible outcomes: first, there could in principal be wormhole solutions sourced by the quantum expectation value of the stress-tensor in the presence of a scalar source at the boundary. While these would be hard to construct, there existence is a logical possibility, even though matter that can support wormholes must have some form of exoticism and the stability of such solutions is typically an issue (see for example \cite{Hebecker:2018ofv} for a review). Second, it is possible that while no on-shell solution connects the two-boundaries, the correlation is still produced in the gravitational path integral by field configurations that connect the two boundaries, which certainly exist. A related investigation will be carried out in \cite{Cotler:2020ugk}.

\section*{Acknowledgements}

We are happy to thank Kristan Jensen, Raghu Mahajan, Henry Maxfield, Gabor Sarosi, Steve Shenker, Julian Sonner, Douglas Stanford, Joaquin Turiaci and Zhenbin Yang for fruitful discussions. A.B. would like to acknowledge the Amsterdam String Workshop, the annual It from Qubit meeting, and the KITP program "Gravitational Holography" where many inspiring discussions for this work took place. The work of A.B. is supported in part by the NWO VENI grant 680-47-464 / 4114. JdB is supported by the European Research Council
under the European Unions Seventh Framework Programme (FP7/2007-2013), ERC Grant
agreement ADG 834878.

\bibliographystyle{unsrt}
\bibliography{ref}
\end{document}